\documentclass[conference]{IEEEtran}
\IEEEoverridecommandlockouts
\usepackage{cite}
\usepackage{amsmath,amssymb,amsfonts}
\usepackage{algorithmic}
\usepackage{graphicx}
\usepackage{textcomp}
\usepackage{xcolor}
\usepackage{fontenc}
\usepackage[procnames]{listings}
\usepackage{lipsum,mwe,cuted}
\usepackage{float}
\usepackage{caption}
\usepackage{booktabs} 
\usepackage{multirow}
\usepackage{tikz}
\usepackage{colortbl}
\usepackage{makecell}
\usepackage{pifont}
\usepackage{multirow}
\usepackage[hidelinks, colorlinks=true, linkcolor=black, urlcolor=blue, citecolor=black]{hyperref}

\def\BibTeX{{\rm B\kern-.05em{\sc i\kern-.025em b}\kern-.08em
    T\kern-.1667em\lower.7ex\hbox{E}\kern-.125emX}}
\begin{document}

\title{Hierarchical Source-to-Post-Route QoR Prediction in High-Level Synthesis with GNNs\vspace{-0.5em}}

\author{\IEEEauthorblockN{Mingzhe Gao$^{1}$,
Jieru Zhao$^{1*}$, Zhe Lin$^{2*}$, Minyi Guo$^{1}$}
\IEEEauthorblockA{${}^{1}$Shanghai Jiao Tong University, ${}^2$Sun Yat-sen University\\
\{a823337391z,zhao-jieru\}@sjtu.edu.cn, linzh235@mail.sysu.edu.cn, guo-my@cs.sjtu.edu.cn}\vspace{-1cm}}

\maketitle
\begingroup\renewcommand\thefootnote{*}
\footnotetext{Jieru Zhao and Zhe Lin are the corresponding authors.}
\endgroup

\begin{abstract}
High-level synthesis (HLS) notably speeds up the hardware design process by avoiding RTL programming. However, the turnaround time of HLS increases significantly when post-route quality of results (QoR) are considered during optimization. To tackle this issue, we propose a hierarchical post-route QoR prediction approach for FPGA HLS, which features: (1) a modeling flow that directly estimates latency and post-route resource usage from C/C++ programs; (2) a graph construction method that effectively represents the control and data flow graph of source code and effects of HLS pragmas; and (3) a hierarchical GNN training and prediction method capable of capturing the impact of loop hierarchies. Experimental results show that our method presents a prediction error of less than 10\% for different types of QoR metrics, which gains tremendous improvement compared with the state-of-the-art GNN methods. By adopting our proposed methodology, the runtime for design space exploration in HLS is shortened to tens of minutes and the achieved ADRS is reduced to 6.91\% on average. Code and models are available at \url{https://github.com/sjtu-zhao-lab/hierarchical-gnn-for-hls}.
\end{abstract}

 \vspace{-0.13cm}   
\section{Introduction}
 \vspace{-0.1cm}  

High-level synthesis (HLS) enables users to create hardware designs at a high level of abstraction using programming languages like C/C++, and provides a series of pragmas, to tune hardware micro-architectures. Due to its superior productivity, HLS is becoming a popular solution for next-generation hardware development, especially for FPGAs. 
However, the turnaround time for HLS design is significantly increased when post-route quality of results (QoR), such as latency and resource utilization, are considered during optimization. To obtain an accurate assessment of post-route QoR metrics, a complete C-to-bitstream design flow should be invoked each time the HLS code is modified.  
This runtime overhead undermines the benefits brought by HLS, especially when it comes to choosing the solution with optimal QoRs from an extremely large design space. 
Therefore, it is essential to provide a fast and accurate QoR prediction framework at an early design stage like HLS to improve design efficiency.

Existing prediction methods can be classified into analytical and learning-based models. Analytical models are constructed based on domain expertise and achieve high accuracy in latency prediction \cite{zhao2017comba, choi17}. However, they are tailored for a specific subset of pragmas, resulting in limited scalability. Learning-based models adopt machine learning techniques \cite{7927161, dai2018fast, makrani2019pyramid} and graph neural networks (GNNs) \cite{sohrabizadeh2022gnn, ferretti2022graph,wu2022high} to predict latency and resource usage in HLS. We compare representative methods in Table \ref{tab:comp-related}. Zhong et al. \cite{7927161} extract features via source code profiling and use the gradient-boosted machine (GBM) to predict post-HLS metrics that can be obtained from HLS reports after synthesis. Then they conduct efficient DSE guided by their model. However, the labels in their dataset, i.e., post-HLS metrics, may deviate significantly from actual post-route QoR values. Hence, predicting post-HLS metrics may mislead the DSE process and get sub-optimal designs. Dai et al. \cite{dai2018fast} and Pyramid \cite{makrani2019pyramid} also utilize ML models while predicting post-route QoR metrics directly. However, their features come from HLS reports and necessitate the execution of the HLS flow, incurring a higher time cost. Recent advances introduce GNN models for QoR prediction due to their representation power of C/C++ programs. GNN-DSE \cite{sohrabizadeh2022gnn} and Ferretti et al. \cite{ferretti2022graph} represent source code as graphs and predict post-HLS metrics, which induce the same issue as \cite{7927161}. Wu et al. \cite{wu2022high} start from the HLS intermediate representation (IR) and predict post-route QoRs. They first predict resource types of operations and then incorporate the predicted resource types as node features to further estimate post-route resource usage. However, their input graphs are constructed on top of HLS IR, requiring running the HLS flow to generate related information. More importantly, the samples in their dataset exhibit relatively simple code structures, including randomly generated DFGs and loops without pragmas applied. The missing pragma modeling makes it less suitable for DSE tasks.    



\begin{table}
\caption{Comparison between representative methods.}
\label{tab:comp-related}
\renewcommand\arraystretch{1}
\centering
\begin{tabular}{cccccc}
\hline
\makecell{} & \makecell{\textbf{Model}} & \makecell{\textbf{Input}} & \makecell{\textbf{Prediction} \\\textbf{Targets}} & \makecell{\textbf{Pragmas} \\ \textbf{Involved}} & \makecell{\textbf{HLS Exec.} \\\textbf{Free}} \\ \hline
\cite{7927161} & GBM & Source code & \makecell{Post-HLS} & \ding{51} & \ding{51} \\
\cite{dai2018fast} & XGB & \makecell{HLS reports} & \makecell{Post-Route} & \ding{55} & \ding{55} \\
\cite{sohrabizadeh2022gnn} & GNN & Source code & \makecell{Post-HLS} &  \ding{51} &  \ding{51}   \\
\cite{wu2022high}  & GNN & HLS IR & \makecell{Post-Route} & \ding{55} &  \ding{55} \\
Ours & GNN & Source code & \makecell{Post-Route} &  \ding{51} & \ding{51}   \\
\hline
\end{tabular}
\vspace{-0.5cm}
\end{table}

To summarize, few of the previous methods achieve accurate source-to-post-route QoR prediction. The problem is even more challenging if the code structure is complicated and multiple HLS pragmas are considered. Different from directly predicting QoR with the whole graph in previous methods, our insight is that we can reserve the structural information of loops/functions and predict post-route QoR metrics gradually from inner to outer hierarchies. This will reduce the estimation difficulty and potentially improve the prediction accuracy. 

To this end, we propose a \textit{hierarchical} prediction method using GNNs, which distinguishes itself from prior arts with the following key features: 1) \textbf{efficiency}: our method directly takes C/C++ source code as input and gives a fast post-route QoR estimation without invoking any EDA design flow, inlcuding HLS, during inference; 2) \textbf{pragma embedding:} we introduce an effective graph construction approach that jointly represents the control and data flow graph of source code and effects of HLS pragmas; 3) \textbf{hierarchical training and prediction}: our method is capable of capturing the structural information of code hierarchies and presents the ability to estimate QoR for different loop/function hierarchies; 4) \textbf{accuracy}: our method presents an average prediction error of $<$10\% for post-route QoR metrics, outperforming the state-of-the-art GNN methods. With efficient and accurate QoR estimation, the DSE time is shortened to tens of minutes and the achieved ADRS is reduced to 6.91\% on average.


\vspace{-0.1cm} 
\section{PRELIMINARIES}
\vspace{-0.1cm} 

\subsection{Control and Data Flow Graph (CDFG)}
\vspace{-0.1cm}




The Control Flow Graph (CFG) is a graphical representation of a program that describes the possible flows between basic blocks in the program. Likewise, the Data Flow Graph (DFG) is another graphical representation that depicts the dependencies between operations. By combining CFG and DFG, we obtain a Control and Data Flow Graph (CDFG). 
Using CDFG to represent programs offers several advantages: 1) CDFG can be generated with LLVM IR, allowing for the earliest possible graph description of a program; 2) CDFG naturally reflects hierarchies of loops, making it convenient to annotate loop-based information at different levels; 3) CDFG contains control and data flow information that can be used to distinguish whether loops can be parallelized and determine whether resources can be shared.


\vspace{-0.1cm} 
\subsection{Graph Neural Network}
\vspace{-0.1cm} 



GNNs process graph-structured data, where objects are nodes and relationships are edges. GNNs learns node representation with a \emph{message passing} mechanism: each node \( v \) in the graph \( G \) collects the embedding \( h^k_{N_v} \) of its \( k \)-hop neighbor \({N_v}\) to update its embedding \( h^k_v \) after \( k \) graph convolution layers.

Since input C/C++ programs can be represented as graphs (i.e., CDFGs) naturally, GNNs become a powerful technique to capture features of each operation and recognize relationships between neighbor operations. Due to its representation power, we make full use of GNNs in a hierarchical way for the post-route QoR prediction.

\begin{figure}
\centering
	 \includegraphics[width=0.45\textwidth]{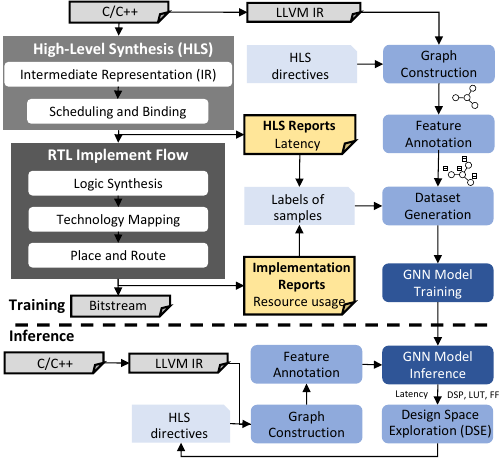} 
	 \caption{Framework overview}
  \vspace{-0.6cm}
\label{fig_overview} \end{figure}

\vspace{-0.1cm} 
\section{Methodology}
\vspace{-0.1cm} 
\label{sec: method}

Figure \ref{fig_overview} illustrates the overview of our proposed approach. 
During the training phase, we deploy the complete C-to-bitstream design flow to obtain the ground-truth QoR considering various combinations of HLS pragmas. We extract resource statistics from implementation reports after place and route (PAR) and latency details from HLS reports. Our approach starts with the input source code (C/C++) which is transformed into LLVM IR. Then we construct a graph representation of the input code with different pragma configurations applied (\textit{graph construction}). After that, node and edge features are extracted and annotated in the graph (\textit{feature annotation}). By bundling the annotated graphs and their corresponding ground-truth labels under the same pragma configuration, our dataset can be generated (\textit{dataset generation}). In the final step, we train multiple GNN models for accurate post-route QoR prediction, while incorporating the inherent hierarchical nature of loops (\textit{GNN model training}). 
During the inference phase, accurate predictions of post-route performance are provided for a given input program, without invoking any HLS or RTL implementation flow. Based on the fast feedback from the prediction model, efficient DSE can then be conducted to search for the optimal configuration of pragmas that achieves the best performance. 

\vspace{-0.1cm}
\subsection{Graph Construction} \label{sec:graph contruction}
\vspace{-0.1cm} 

The input source code is first transformed to LLVM IR using the Clang front-end. Then we construct the corresponding graph representation by extending a CDFG generator, PrograML \cite{cummins2021a}, which captures both control and data flow. However, PrograML does not target FPGAs, and the generated program graphs need to be extended to consider the effects of HLS pragmas. Figure \ref{fig_graph} illustrates our graph construction process considering different kinds of pragmas.

\subsubsection{Constructing graph with loop pipelining} In a pipelined loop, the execution of consecutive iterations can be initiated after a specific interval, i.e., initiation interval (II). This is achieved by inserting registers into the logic during the synthesis process of HLS tools. The equivalent graph representation is not influenced or changed at the compilation stage. Therefore, the graphs constructed for pipelined loops remain the same as those of original loops without pipelining, as shown in Fig. \ref{fig_graph}(a). 

\subsubsection{Constructing graph with loop unrolling} 
The loop unrolling pragma facilitates the concurrent execution of iterations and results in the creation of multiple replicas of hardware logic. In accordance with this principle, when constructing graphs for unrolled loops, the pertinent logic nodes in the unrolled region are replicated and appropriate edges are established to connect to original predecessors and successors. This procedure is illustrated in Fig. \ref{fig_graph}(b). By adding replicas of logic nodes explicitly, the resulting graph achieves a natural representation of \textit{loop unrolling} and significantly alleviates the complexity of prediction.

\subsubsection{Constructing graph with array partitioning} 
\begin{figure}
\vspace{-0.2cm}
        \centering
	 \includegraphics[width=3.5in]{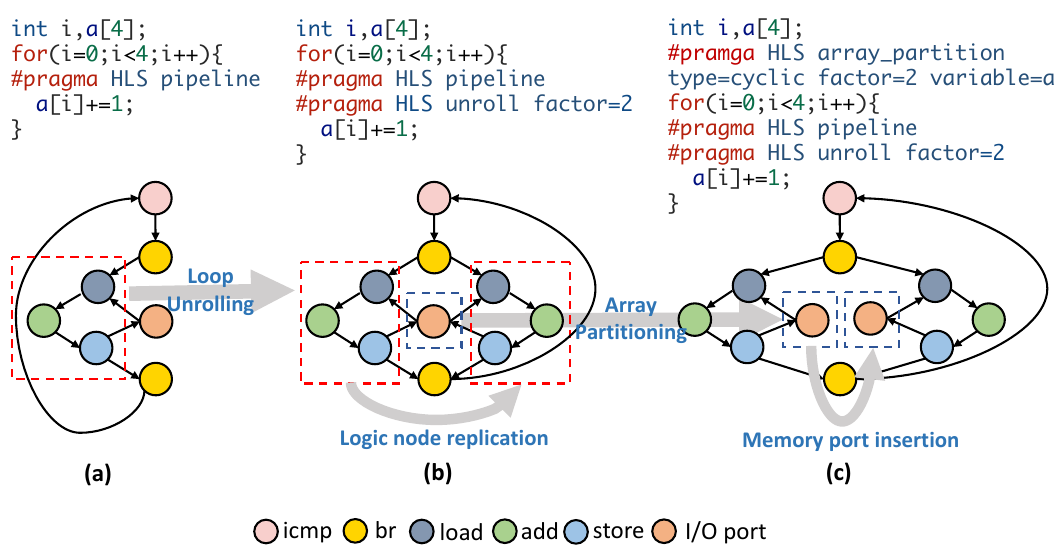} 
	 \caption{Procedure of graph construction.}
  \vspace{-0.5cm}
\label{fig_graph} \end{figure}

HLS tools provide the \textit{array partitioning} pragma to increase local memory bandwidth. Arrays are divided into smaller ones based on user-specified options. To reflect the impact of memory ports on performance estimation, we incorporate I/O port nodes to the original CDFG, as highlighted in orange in Fig. \ref{fig_graph}. When \textit{array partitioning} is applied, the corresponding memory port nodes are split into a certain number of nodes based on partitioning factors. These newly introduced memory port nodes are connected to load and store nodes that read from or write to corresponding memory locations. 
The connections are determined by partitioning types as well as the memory access pattern of the code. Specifically, for a $k$-dimensional array, assuming its partition factors across the $k$ dimensions are $u_1, u_2, \cdots, u_k$, we add $\prod_{i=1}^{k} u_i$ memory port nodes in the graph. Simultaneously, we write LLVM passes to analyze the index values of each load and store operation to determine which memory ports should be connected to them.  If the array index is dynamic or unpredictable, the load/store nodes will connect to all memory ports.

\vspace{-0.1cm}
\subsection{Feature Extraction and Annotation}
\label{ssec: feat}
\vspace{-0.1cm} 

Effective features need to be extracted from the code and annotated to the generated graph for model training. We categorize the features we have used into two categories: (1) node features and (2) loop-level features. 

\subsubsection{Node Features} Table \ref{tab-feature} summarizes our node features.
\textbf{Optype} denotes the operation type of each node, such as \verb|add|, \verb|mul|, \verb|div|, \verb|load|, \verb|store|. This is obtained by analyzing LLVM IR instructions. \textbf{\#invocation} denotes the number of times that an operation or a node is executed in a loop, which is associated with the loop tripcount and unroll factor. 
\textbf{In and out degrees} represent the number of edges flowing into and out of the node, respectively, which can indirectly reflect the potential resource usage of multiplexers.
\textbf{Delay} and \textbf{\#cycle} represent the timing information of each operation/node, which helps latency prediction. We build a latency and delay library for each operation type based on profiling results from several micro-benchmarks.
For \textbf{LUT}, \textbf{DSP} and \textbf{FF}, we profile corresponding resource usage for arithmetic operations (e.g., \verb|add|, \verb|mul|, \verb|fadd|) using micro-benchmarks and build a library. For non-arithmetic operations (e.g., \verb|br|, \verb|icmp|, \verb|mux|), we simply set their resource-related features to zero.

\begin{table}
\renewcommand\arraystretch{1.1}
\caption{Node features}
\vspace{-0.2cm}
\begin{center}
\setlength{\tabcolsep}{2mm}{
\begin{tabular}{cccc}
\hline
\textbf{Feature} & \textbf{Description} & \textbf{Values} \\
\hline
Optype    & operation type  & \verb|load|, \verb|etc.| \\
\#invocation     & the number of invocations & \verb|int| \\
In degree   & the number of incoming edge  & \verb|int| \\
Out degree    & the number of outcoming edge & \verb|int| \\
\#cycle  & the number of clock cycles  & \verb|int| \\
Delay  & the delay (ns) of each operation  & \verb|float| \\
LUT    & LUT usage of each operation & \verb|int| \\
DSP    & DSP usage of each operation & \verb|int| \\
FF   & FF usage of each operation & \verb|int| \\
\hline
\end{tabular}}
\label{tab-feature}
\end{center}
\vspace{-0.5cm}
\end{table}

\begin{figure*}
        \centering
	 \includegraphics[width=0.95\linewidth]{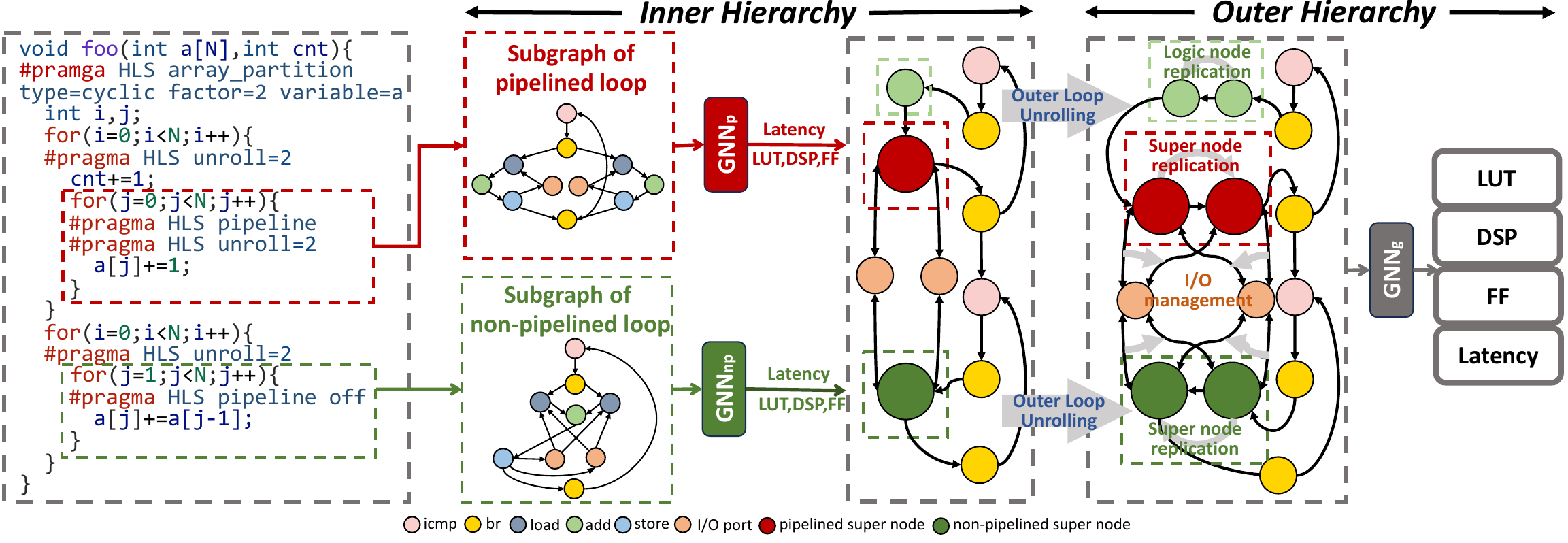} 
	 \caption{The illustration of our hierarchical source-to-post-route QoR modeling approach.}
  \vspace{-0.4cm}
\label{fig_hierarchyGNN} \end{figure*}


\subsubsection{Loop-level features} Besides node features, we extract loop-level features so as to consider the effect of HLS pragmas on loops. As discussed in Section \ref{sec:graph contruction}, the graph of a pipelined loop remains the same as that of the loop without pipelining. Therefore, additional features should be considered to differentiate pipelining and non-pipelining. We use three features that determine the latency of a pipelined loop, namely, iteration latency (\textit{IL}), initiation interval (\textit{II}) and tripcount (\textit{TC}). Specifically, \textit{IL} can be approximated by GNNs and \textit{TC} can be obtained by analyzing the LLVM IR code, while II can be computed by the equation\cite{8695879}:
\vspace{-0.1cm}
\begin{eqnarray*}
II_\textit{min} & = & \max(II^{rec}, II^{res}) \\
   & = & \max(\max\{\lceil \frac{\textit{Delay}_p}{\textit{Distance}_p} \rceil\}, \max\{\lceil \frac{\textit{Access}_m}{\textit{Ports}_m} \rceil\})
\end{eqnarray*}
where Delay$_p$ represents the latency between a pair of dependent instructions from different iterations, Distance$_p$ refers to the difference between the corresponding iteration numbers, Ports$_m$ is the number of ports and Access$_m$ is the number of accesses to array $m$. 

\vspace{-0.1cm} 
\subsection{The Hierarchical Modeling Approach}
\label{ssec: model}
\vspace{-0.1cm} 

After constructing and annotating graphs, the complete C-to-bitstream flow is invoked to obtain the ground truth QoR considering various HLS pragmas. Then the annotated graphs and their corresponding QoR labels form our dataset for model training. The detailed setup is discussed in Section \ref{sec:experiment setup}. 

We propose a hierarchical source-to-post-route QoR modeling methodology, as depicted in Fig. \ref{fig_hierarchyGNN}. To capture and incorporate the inherent hierarchical nature of loops, we train multiple GNN models to estimate QoR results at different loop hierarchies and gradually predict the overall latency and resource usage of the application. Our method can be generalized to applications with multiple nested loops. Take the code in Fig. \ref{fig_hierarchyGNN} as an example, the top function \textit{foo} has two nested loops, each with two loop levels, i.e., loop hierarchies. In the first nested loop, its sub-loop at the inner hierarchy (\textit{j-level}) is pipelined and partially unrolled, and its sub-loop at the outer hierarchy (\textit{i-level}) is partially unrolled with factor two. This configuration creates a circuit with two replicas of the inner loop pipeline within the outer loop. The QoR of inner replicas greatly impacts the overall QoR of the nested loop. Thus, our strategy models the QoR of sub-loops at the inner hierarchy using a local GNN, merges inner sub-loops to \textit{super nodes}, and integrates super nodes with operations at the outer hierarchy 
to model overall QoR with a global GNN.


This hierarchical approach simplifies post-route QoR prediction by dividing the problem into smaller ones. It also improves estimation accuracy by capturing fine-grained features at different hierarchies, compared to previous GNN methods that directly predict performance with the whole graph. To be more specific, we divide loop hierarchies into two levels: \textbf{inner and outer hierarchies}. 

\subsubsection{Inner Hierarchy}
In our method, a loop at the \textbf{inner hierarchy} denotes the loop that only contains computing logic without sub-loops. There are four types of loops that belong to the inner hierarchy: \ding{172} a single-level loop; \ding{173} a nested loop with \textit{loop pipelining} applied to the outermost level (in this case, all the inner sub-loops will be fully unrolled); \ding{174} a perfect nested loop with \textit{loop flattening} and \textit{loop pipelining} applied to the innermost level (in this case, all the loop levels will be flattened to form a deeper pipeline which is equivalent to a single-level pipelined loop); and \ding{175} a nested loop with all the inner sub-loops fully unrolled. For example, in Fig. \ref{fig_hierarchyGNN}, the two j-level loops in red and green boxes are loops at the inner hierarchy, corresponding to category \ding{173} and \ding{172}, respectively.
At the inner hierarchy, related loops are first extracted and subgraphs are constructed for model training, as illustrated in Fig. \ref{fig_hierarchyGNN}. Then we train two different GNN models, namely $\textbf{GNN}_p$ and $\textbf{GNN}_{np}$, to predict QoR for pipelined and non-pipeline loops, respectively. This is because execution models of pipelined and non-pipelined loops are different and training GNN models separately can improve accuracy based on our experiments. After predicting the QoR of loops at the inner hierarchy, the nodes in a subgraph 
are merged into a super node, which represents this loop at the inner hierarchy when modeling the outer hierarchy. 

\subsubsection{Outer Hierarchy} We define all the remaining logic and loop levels as components at the \textbf{outer hierarchy}. Note that loops can contain more than two loop levels. After the first stage, we condense each loop at the inner hierarchy into a super node. These super nodes are connected to the nodes from the outer hierarchy to form the complete graph representation of the top function \textit{foo}. If the outer loop is unrolled, corresponding logic and super nodes are replicated following strategies discussed in Section \ref{sec:graph contruction}. In the entire program graph, we 
annotate super nodes with predicted QoR as features. Each super node holds a complete set of node features as shown in Table~\ref{tab-feature}, where features such as \verb|LUT|, \verb|DSP|, \verb|FF|, \verb|latency| are derived from the previous prediction phase. The calculation method used for other features remains consistent with those of standard logic nodes. This program graph, containing super nodes and logic nodes of the outer hierarchy, becomes the input graph of our global GNN model named $\textbf{GNN}_g$. The global GNN model is trained to predict the final QoR of the whole application. This approach not only assists us in predicting the overall performance of real applications more accurately but also helps in discerning serial and parallel relationships between super nodes, i.e., loops.

\begin{figure*}
    \centering
	 \includegraphics[width=0.95\linewidth]{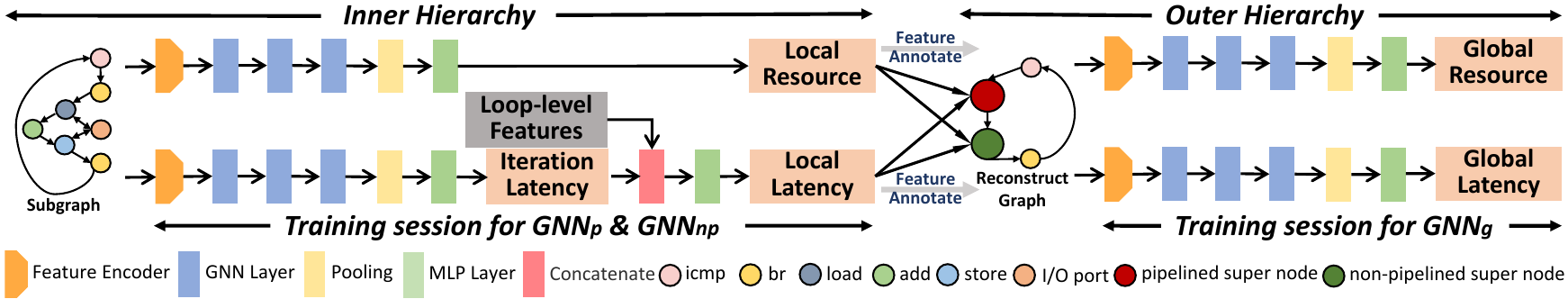} 
	 \caption{GNN architectures and the corresponding hierarchical training process.}
\label{fig_GNNarch} 
\vspace{-0.3cm}
\end{figure*}

\vspace{-0.15cm} 
\subsection{GNN Architectures and Training Process} 
\label{gnnarch}
\vspace{-0.1cm} 

We develop three GNN models for the prediction of each QoR metric, namely $\textbf{GNN}_p$, $\textbf{GNN}_{np}$, and $\textbf{GNN}_g$. 
Due to different execution models on hardware, $\textbf{GNN}_p$ and $\textbf{GNN}_{np}$ are deployed for pipelined and non-pipelined loops at the inner hierarchy, respectively. $\textbf{GNN}_g$ is used for prediction at the outer hierarchy. Figure \ref{fig_GNNarch} illustrates the model architectures and our hierarchical training process. 
The training process of the three GNN models is carried out in a hierarchical way. We initially collect the loops at the inner hierarchy, categorize them into pipelined loops and non-pipelined loops, and extract information from the two kinds of loops to generate corresponding datasets. Then $\textbf{GNN}_p$ and $\textbf{GNN}_{np}$ are trained accordingly. Once the training process is completed, the weights of both $\textbf{GNN}_p$ and $\textbf{GNN}_{np}$ are frozen. The output of these two GNNs is then used as features of super nodes, which are combined with features of other operations at the outer hierarchy to generate the dataset for training $\textbf{GNN}_g$.


Our GNN models have similar architectures, as depicted in Fig. \ref{fig_GNNarch}. Each model can be decomposed into 1) feature encoder, 2) propagation layer, 3) pooling layer, and 4) MLP layer.  The \textit{feature encoder} converts the attributes of nodes into initial feature vectors. We use one-hot encoding for the attribute \verb|optype|, while numerical attributes are directly incorporated into the feature vector. \textit{Propagation layers} implement the message-passing mechanism of GNNs. We adopt the propagation layers from typical GNN models: GCN~\cite{kipf2017semi}, GAT~\cite{velickovic2018graph}, GraphSAGE~\cite{hamilton2017inductive}, TransformerConv~\cite{shi2020masked}, and PNA~\cite{Corso2020PrincipalNA}.
After passing through three propagation layers, node embeddings are obtained. 
Then we use a \textit{pooling layer} to generate graph-level embeddings from node embeddings. We use two types of pooling: sum pooling \(h_{sum}\) and max pooling \(h_{max}\). The embeddings from these two pooling mechanisms are concatenated to form the graph-level representation.
Finally, \textit{MLP layers} are used to project the graph-level embeddings into QoR prediction. For resource usage, we directly obtain the estimates after feeding the embeddings of the GNN into an MLP layer. For latency prediction, we handle the inner hierarchy ($\textbf{GNN}_p$ and $\textbf{GNN}_{np}$) and the outer hierarchy ($\textbf{GNN}_{g}$) separately. Regarding the inner hierarchy, first, $\textbf{GNN}_p$ and $\textbf{GNN}_{np}$ utilize an MLP layer to predict iteration latency. Second, another MLP layer takes the predicted iteration latency and loop-level features (elaborated in Section~\ref{ssec: feat}) as input and infers the latency of the loops at the inner hierarchy. Regarding the outer hierarchy, we directly predict the overall latency with an MLP.

\vspace{-0.15cm}
\section{Experiments}
\vspace{-0.12cm}

\subsection{Experimental setup} \label{sec:experiment setup}
\vspace{-0.12cm} 

Our feature construction and model development flow is fully automated and implemented using Python and C++. We utilize 16 applications from the Polybench, MachSuite and CHStone benchmark suite. Specifically, 12 applications are used for GNN model training and testing, and 4 applications are used for the DSE experiment. The datasets to develop $\textbf{GNN}_p$ and $\textbf{GNN}_{np}$ are constructed based on sub-loops extracted from the application source code, while the dataset for $\textbf{GNN}_g$ is constructed based on the complete applications. Our experiments are conducted on the AMD Ultrascale+ MPSoC ZCU102. 
We utilize Vitis-HLS 2022.1 and Vivado 2022.1 to collect ground-truth latency and resource usage as training labels. By applying various pragma combinations, we produce 3102, 2300 and 6178 valid designs to build \(\textbf{GNN}_p\), \(\textbf{GNN}_{np}\), and \(\textbf{GNN}_g\), respectively. 80\% of the dataset is used for training, 10\% of the dataset is used for validation, and the remaining 10\% of the dataset is used for testing. We use the mean absolute percentage error (MAPE) to quantify the prediction quality. Each model is trained with over 250 epochs.

\begin{table}
\caption{MAPE of post-route QoR with different GNNs.}
\label{tab:predication}
\centering
\setlength{\tabcolsep}{1mm}{
\begin{tabular}{c|cccccc}
\hline
\textbf{GNN type} &    & \textbf{Latency}  & \begin{tabular}[c]{@{}c@{}}\textbf{Iteration}\\ \textbf{Latency }\end{tabular}& \textbf{DSP}  & \textbf{LUT}  & \textbf{FF}  \\ \hline
\multirow{3}{*}{GCN}  
& $\textbf{GNN}_p$      & 5.23\%     & 7.21\% & 4.94\% & 6.63\%  & 7.83\%  \\
& $\textbf{GNN}_{np}$       & 8.41\% & 8.41\% & 9.35\% & 5.88\%  & 8.39\% \\
& \cellcolor{lightgray} $\textbf{GNN}_g$      & \cellcolor{lightgray}9.85\% &\cellcolor{lightgray} N/A & \cellcolor{lightgray}7.45\% & \cellcolor{lightgray}11.27\% & \cellcolor{lightgray}11.24\% \\
\hline
\multirow{3}{*}{GAT}  
& $\textbf{GNN}_p$       & 4.91\% & 7.89\%  & 4.67\%  & 9.12\% & 9.20\%  \\
& $\textbf{GNN}_{np}$       & 9.42\% & 9.42\%  & 10.64\%  & 6.99\% & 9.32\%  \\
&\cellcolor{lightgray} $\textbf{GNN}_g$      & \cellcolor{lightgray} 10.88\% & \cellcolor{lightgray} N/A & \cellcolor{lightgray} 8.18\%  &  \cellcolor{lightgray}12.34\%  & \cellcolor{lightgray}11.57\% \\
\hline
\multirow{3}{*}{GraphSage}
& $\textbf{GNN}_p$ & 5.24\% & 5.52\%   & 5.88\%  & 8.12\% & 8.66\%   \\
& $\textbf{GNN}_{np}$ & 7.47\% & 7.47\% & 9.83\% & 6.25\% & 7.22\%   \\
& \cellcolor{lightgray}$\textbf{GNN}_g$ &\cellcolor{lightgray} 8.55\% &\cellcolor{lightgray} N/A & \cellcolor{lightgray}\textbf{6.94\%} &\cellcolor{lightgray} 9.86\% & \cellcolor{lightgray} \textbf{9.99\%} \\
\hline
\multirow{3}{*}{Transformer}
& $\textbf{GNN}_p$ & 4.92\% & 5.95\% & 4.05\% & 6.70\% & 7.12\% \\
& $\textbf{GNN}_{np}$ & 8.77\% & 8.77\%  & 12.02\%  & 5.83\% & 8.65\% \\
& \cellcolor{lightgray}$\textbf{GNN}_g$ & \cellcolor{lightgray}\textbf{8.54\%} & \cellcolor{lightgray}N/A & \cellcolor{lightgray} 6.97\% &  \cellcolor{lightgray}11.33\% & \cellcolor{lightgray} 10.31\%  \\
\hline
\multirow{3}{*}{PNA}
& $\textbf{GNN}_p$       & 6.74\% & 7.43\% & 4.11\%  & 8.92\%  & 9.93\%  \\ 
& $\textbf{GNN}_{np}$       & 9.99\% & 9.99\% & 10.55\% & 7.89\% & 10.26\%  \\ 
& \cellcolor{lightgray}$\textbf{GNN}_g$       & \cellcolor{lightgray}9.29\% &\cellcolor{lightgray} N/A &\cellcolor{lightgray} 7.84\% & \cellcolor{lightgray}\textbf{9.65\%} &\cellcolor{lightgray} 10.55\%  \\
\hline
\end{tabular}}
\vspace{-0.4cm}
\end{table}

\vspace{-0.1cm} 
\subsection{Evaluation of QoR prediction accuracy}
\vspace{-0.1cm} 

As introduced in Section \ref{gnnarch}, we adopt five different GNN models and compare their QoR prediction accuracy, as shown in Table \ref{tab:predication}. For loop-level models ($\textbf{GNN}_p$ and $\textbf{GNN}_{np}$), we present prediction errors for both latency and iteration latency. For application-level models ($\textbf{GNN}_g$), the latency prediction error of complete designs is reported. Among all the GNN models for application-level prediction, Transformer, GraphSage, and PNA achieve the lowest estimation errors in latency (8.54\%), DSP usage (6.94\%) and FF usage (9.99\%), and LUT usage (9.65\%), respectively. Our hierarchical method leads to high prediction accuracy of post-route QoR for pipelined loops ($\textbf{GNN}_p$), non-pipelined loops ($\textbf{GNN}_{np}$) and the complete applications ($\textbf{GNN}_g$), showing prediction errors of less than 10\% for latency and all types of resources. Since our method directly takes the C/C++ source code as input to predict the post-route QoR of FPGA designs without invoking any EDA tool flow, the prediction accuracy is highly desirable.

\vspace{-0.1cm} 
\subsection{Comparison with the state-of-the-art}
\vspace{-0.1cm} 

We compare our method with the approach \cite{wu2022high} which also predicts post-route QoR using GNNs. Different from\cite{wu2022high}, we additionally consider various pragma combinations, increasing the complexity of code structures. For a fair comparison, we evaluate both methods on the dataset from \cite{wu2022high} which does not consider HLS pragmas and our dataset with pragmas, as shown in Table \ref{tab:comparison}. Without pragmas, our method and \cite{wu2022high} achieve close prediction accuracy, while on the dataset with pragmas applied, our method demonstrates superior improvement in all QoR metrics. We attribute this notable improvement to the following two reasons: 1) \textit{joint representation of code and pragmas during graph construction}: we deploy a graph construction method that represents the CDFG of source code while taking into account the impact of pragmas. This approach effectively embeds the pragma information into the graph representation. 2) \textit{reservation of loop hierarchies}: with our hierarchical training approach, messages can propagate more effectively between different levels of loop hierarchies, allowing GNNs to better understand the cascading effects of changes in the design's inner hierarchy on the outer hierarchy.

\begin{table}
\caption{Comparison of prediction error (MAPE).}
\vspace{-0.2cm}
\label{tab:comparison}
\begin{center}
\begin{tabular}{cccccc}
\hline
\textbf{Method} & \textbf{Configuration} & \textbf{Latency} & \textbf{DSP} & \textbf{LUT} & \textbf{FF} \\ \hline
\cite{wu2022high}  & w/o pragma & N/A & 8.26\% & 5.10\% & 7.58\%  \\
Ours & w/o pragma & 5.54\% & 6.71\% & 6.78\% & 7.91\% \\
\hline
\cite{wu2022high}  & w/ pragma & 35.81\% & 57.31\% & 27.14\% & 29.03\% \\
Ours & w/ pragma & 8.54\% & 6.94\% & 9.65\% & 9.99\% \\     
\hline
\end{tabular}
\label{tab1}
\end{center}
\vspace{-0.5cm}
\end{table}



\vspace{-0.2cm} 
\subsection{Evaluation of Design Space Exploration}
\vspace{-0.1cm} 


To demonstrate the effectiveness of our hierarchical QoR prediction method, we perform DSE on four new applications (\verb|bicg|, \verb|symm|,  \verb|mvt| and \verb|syrk|) that are unseen by our model during training. The design space of each application is constructed by varying pragma configurations. Specifically, we apply HLS pragmas like \textit{loop pipelining}, \textit{loop flattening} and \textit{loop unrolling} iteratively from inner to outer loops, with unroll factors from \{1, 2, 4, 8, 16\}. As for \textit{array partitioning}, we set partitioning factors consistent with unroll factors. 

The objective of DSE is to find the optimal pragma configurations producing the best QoR metrics, which form the Pareto frontier. To quantify the optimality, the average distance from reference set (ADRS) can be computed:
\vspace{-0.1cm}
\begin{equation*}
\text{ADRS}(\Gamma, \Omega) = \frac{1}{|\Gamma|} \sum_{\gamma \in \Gamma} \min_{\omega \in \Omega} f(\gamma, \omega),
\end{equation*}

where $\Gamma$ is the exact Pareto-optimal set, $\Omega$ is the approximate Pareto-optimal set, and $f(\cdot)$ computes the distance between two design points. A lower ADRS means that the approximate Pareto set is closer to the exact one. 

To get the exact Pareto-optimal set, we exhaustively evaluate the actual post-route QoRs for each design configuration by invoking the complete C-to-bitstream flow. Vitis HLS and Vivado are utilized to generate the accurate results. We then select the models with the highest prediction accuracy in Table \ref{tab:predication} to provide quick QoR estimation for DSE. For comparison, two other GNN models from \cite{wu2022high} and GNN-DSE~\cite{sohrabizadeh2022gnn} are tested similarly on the same design space. GNN-DSE~\cite{sohrabizadeh2022gnn} is one of the state-of-the-art works in HLS DSE. It uses a GNN-based post-HLS QoR prediction model to address the DSE problem.
The comparison of DSE results is shown in Table \ref{tab:dse}. 
The design space contains up to 2796 design configurations. The DSE time shows the timing costs required to finish the exploration with the QoR feedback from Vivado and from our model, respectively. By utilizing our method, efficient QoR feedback is provided, only necessitating constructing an IR graph from the C-level code followed by several rounds of GNN inference. This effectively bypasses the tedious steps of synthesis and implementation and shortens the DSE time from tens of days to tens of minutes.

Compared to the work~\cite{wu2022high} and GNN-DSE~\cite{sohrabizadeh2022gnn}, which achieve average ADRS values of 13.94\% and 10.96\%  respectively, our method reaches an average ADRS of 6.91\%, demonstrating our superior ability to enable a high-quality DSE. The result fully verifies that the high accuracy of our predictive model can lead to a more precise approximation to the exact Pareto frontier. Regarding the runtime, both GNN-DSE and our method complete the exploration within minutes, while the method\cite{wu2022high} requires one to two days to complete due to the invocation of HLS tools. In summary, by incorporating our hierarchical prediction method, fast and accurate performance feedback can be provided to boost the efficiency of DSE.

\begin{table}
\caption{DSE results on unseen applications.}
\vspace{-0.2cm}
\label{tab:dse}
\centering
\begin{tabular}{c|c|cc|ccc}
\hline
\multirow{2}{*}{\textbf{Kernels}} & \multirow{2}{*}{\textbf{\makecell{\#Design \\ Config}}} & \multicolumn{2}{c|}{\textbf{DSE time}} & \multicolumn{3}{c}{\textbf{ADRS(\%)}} \\ \cline{3-7} 
                         &                                  & \textbf{Vivado}        & \textbf{Ours}        & \textbf{\cite{wu2022high}}      & \textbf{\cite{sohrabizadeh2022gnn}}      & \textbf{Ours}      \\ \hline
bicg                     &    2796                          &  26 days           &  12 min        & 12.13       &  9.14      &  6.39      \\
symm                     &    1972                          &  44 days           &  15 min        & 13.15       &  11.50      &  5.45      \\
mvt                      &    2116                          &  25 days           &  14 min       &  16.01      &    12.17    &   9.31     \\
syrk                     &    1972                          &  40 days           &  15 min        &  14.45      &  11.03      &  6.49      \\ \hline
\end{tabular}
\label{dseruntime}
\vspace{-4mm}
\end{table}

\vspace{-0.15cm}
\section{Conclusion}
\vspace{-0.1cm}


In this paper, we introduce a GNN-based hierarchical method for FPGA post-route QoR modeling from C-level source code.
This efficient source-to-post-route QoR prediction strategy notably reduces the design turn-around time of FPGA HLS. Specifically, we utilize loop hierarchies during model establishment to achieve better prediction accuracy and propose an effective graph construction method to jointly represent code and HLS pragmas. Experiments show that our method outperforms existing works in the prediction of latency and resource usage, which can better support the DSE task.

\vspace{-0.1cm} 
\section*{Acknowledgement}
\vspace{-0.1cm} 
This work is partially sponsored by the National Natural Science Foundation of China (62102249, 62232015) and Shanghai Pujiang Program (21PJ1408200). Mingzhe Gao and Jieru Zhao contributed equally to this paper.

\vspace{-0.1cm} 
\bibliographystyle{IEEEtran}
\bibliography{ref}

\end{document}